\documentclass[conference]{IEEEtran}
\IEEEoverridecommandlockouts
\usepackage{cite}
\usepackage{amsmath,amssymb,amsfonts}
\usepackage{algorithmic}
\usepackage{graphicx}
\usepackage{textcomp}
\usepackage{xcolor}

\usepackage{cite}
\usepackage{epsfig}
\usepackage[center]{caption}
\usepackage{tabularx,booktabs}
\usepackage{comment}
\usepackage[hidelinks]{hyperref}
\usepackage{soul}  
\usepackage[table,xcdraw]{xcolor}
\usepackage{multirow}
\usepackage{svg}
\usepackage{multicol}
\usepackage{makecell} 
\usepackage{enumitem}

\begin{document}

\title{Ising-ReRAM: A Low Power Ising Machine ReRAM Crossbar for NP Problems}

\author{\IEEEauthorblockN{Everest Bloomer, Irem Didin, Ching-Yi Lin and Sahil~Shah}

\IEEEauthorblockA{Department of Electrical and Computer Engineering \\ University of Maryland, College Park, MD, USA. \\e-mail: ebloome1@umd.edu, sshah389@umd.edu.}
}

\maketitle

\begin{abstract}
Computational workloads are growing exponentially, driving power consumption to unsustainable levels. Efficiently distributing large-scale networks is an NP-Complete problem equivalent to Boolean satisfiability (SAT), making it one of the core challenges in modern computation. To address this, physics and device inspired methods such as Ising systems have been explored for solving SAT more efficiently. In this work, we implement an Ising model equivalence of the 3-SAT problem using a ReRAM crossbar fabricated in the Skywater 130 nm CMOS process. Our ReRAM-based algorithm achieves $91.0\%$ accuracy in matrix representation across iterative reprogramming cycles. Additionally, we establish a foundational energy profile by measuring the energy costs of small sub-matrix structures within the problem space, demonstrating under linear growth trajectory for combining sub-matrices into larger problems. These results demonstrate a promising platform for developing scalable architectures to accelerate NP-Complete problem solving.
\end{abstract}


\section{Introduction}
Computational workloads are growing rapidly alongside rising demand for intelligent services, automation, and large-scale data processing, with machine learning (ML) usage for business nearly doubling between Fall 2023 and Fall 2024 \cite{bonney2024tracking}. This growth has driven unprecedented energy consumption, as training large models and managing distributed workloads consume terawatt-hours annually \cite{jouppi2021ten, ahsan2025higher}. Achieving practical throughput increasingly requires distributing computation across many processors, yet optimally scheduling such distributed algorithms is NP-Complete \cite{ullman1975np}. NP-Complete problems lie at the core of challenges in scheduling, optimization, electronic design automation, and neural architecture search, where even modest solution improvements can yield significant gains in energy efficiency and scalability. Consequently, non-conventional computing paradigms such as quantum annealing and Ising-based hardware have emerged as promising accelerators for combinatorial optimization \cite{si2024energy}. Among these, the 3-SAT problem serves as a canonical NP-Complete benchmark, making it an ideal testbed for evaluating hardware-accelerated and physics-inspired computational models.

\begin{figure}
    \centering
    \includegraphics[width=0.85\linewidth]{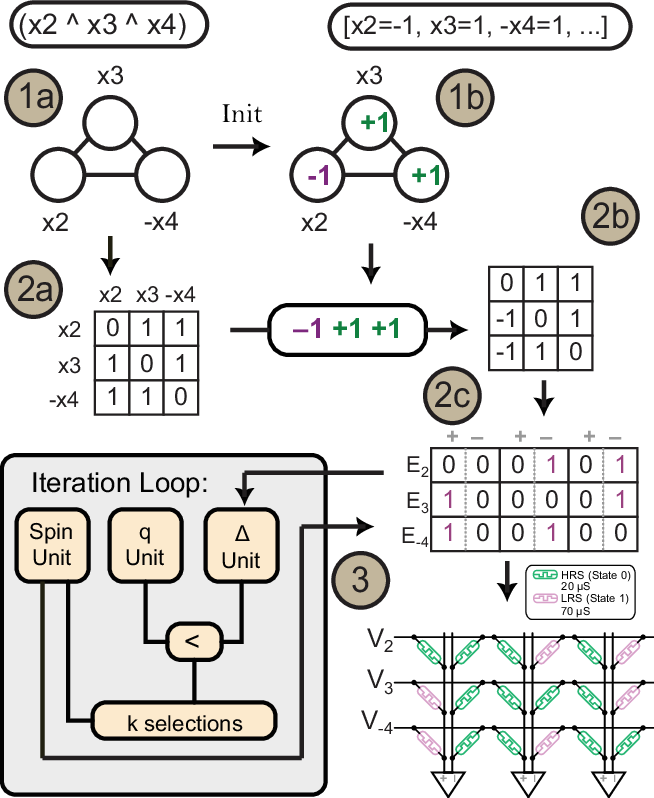}
    \caption{Flowchart overall process, python and hardware. The numbers correspond to the the algorithm flow discussed in Subsection \ref{SubSec:Algo}}
    \label{fig1:flowchart}
    \vspace{-10mm}
\end{figure}

The Ising model, which describes magnetic spin interactions, provides a physical analog for combinatorial optimization by evolving toward a ground state that minimizes an energy function equivalent to a problem’s cost function. NP-Complete problems can be mapped directly onto an Ising graph, where each literal in a 3-SAT instance corresponds to a node and edges encode clause relationships or logical negations \cite{10.3389/fphy.2014.00005}. Under this formulation, efficiently solving the Ising model is equivalent to solving the encoded NP-Complete problem. However, directly simulating magnetic Ising dynamics results in chaotic differential equations \cite{mohseni2022ising}, necessitating tractable digital approximations such as the Metropolis algorithm \cite{metropolis1953equation}. Recently, these approaches have been extended to non-volatile hardware platforms including ReRAM-based systems \cite{chiang2024reaim, pedretti2025solving} and mixed-signal circuit implementations \cite{mathews2024architectural}. Mapping large scheduling and optimization problems onto Compute-in-Memory (CIM) architectures both reduces the cost of memory access and accelerates computation.

In this study, we develop and implement a 3-SAT solving algorithm on a ReRAM crossbar fabricated in the Skywater 130 nm CMOS process. ReRAMs are CMOS-compatible non-volatile devices with relatively low write energy \cite{Didin_2025}. By transforming the 3-SAT problem into a matrix equivalence, we encode it within the crossbar to profile the energy consumption of small instances and analyze the recurring sub-matrix structures relevant to scaling. We further evaluate ReRAM switching accuracy and variability as key factors influencing iterative convergence and computational efficiency.

\section{Methods}

\begin{figure}
    \centering
    \includegraphics[width=0.8\linewidth]{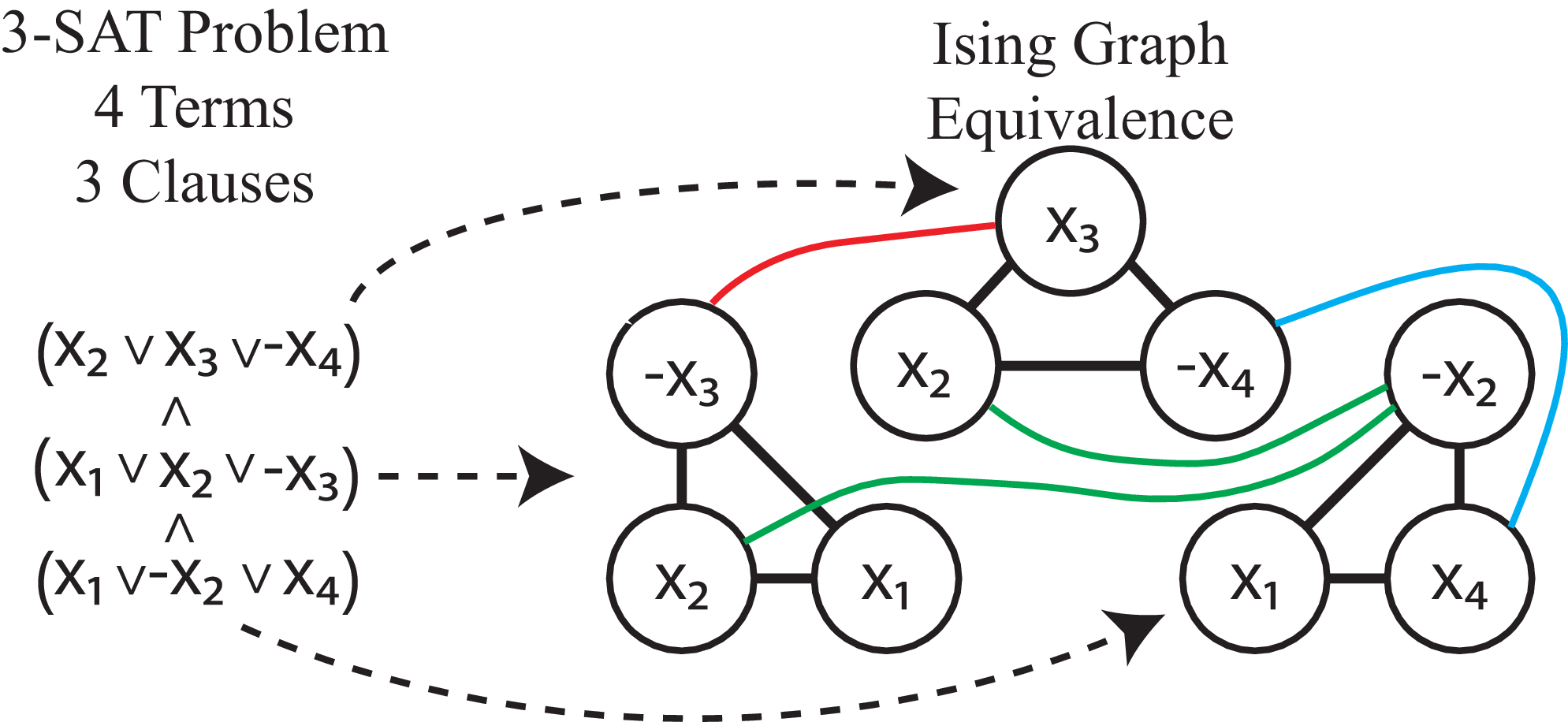}
    \caption{Illustration of 3-SAT to an Ising Graph.}
    \label{fig:graph_build}
    \vspace{-9mm}
\end{figure}

Our process is divided into two components. First, the ReRAM array on-chip \cite{Didin_2025}, which performs the key physical operations of the algorithm as Matrix Multiplication. The second component is the algorithmic framework, implemented through a Python-based simulator that encapsulates all logic executed outside the ReRAM. The complete process is illustrated in Fig. \ref{fig1:flowchart}, with labels for matching sections. 

\subsection{Algorithm} \label{SubSec:Algo}

\begin{enumerate}[leftmargin=1em]
    \item \textbf{Problem Initialization:} \\ (Fig. \ref{fig1:flowchart}, 1a) Begin with the 3-SAT problem in Conjunctive Normal Form (CNF) and transform it into its equivalent graphical form, following the direct transformation method described in \cite{10.3389/fphy.2014.00005} and illustrated in Fig. \ref{fig:graph_build}. (Fig. \ref{fig1:flowchart}, 1b) Assign a random initial spin state to each node in the graph. 
    \item \textbf{Crossbar Mapping:} \\
        (Fig. \ref{fig1:flowchart}, 2a) Convert the graph into its adjacency matrix form. (Fig. \ref{fig1:flowchart}, 2b) Assign the sign of each column according to the initial spin of the corresponding node. Specifically, if node $i$ has spin $-1$, then all entries in column $i$ of the adjacency matrix are negated. \\ (Fig. \ref{fig1:flowchart}, 2c) \textit{Column Expansion and Sign Reversal}: For each column $i$, split it into two columns to represent positive and negative weights in the $ReRAM$. Write the cells pairwise opposite of the value in the adjacency matrix. If the adjacency entry $Adj(i, j) = -1$, then $ReRAM(i, 2j) = 1$ and $ReRAM(i, 2j + 1) = 0$.
    \item \textbf{Main Iterative Loop}: (Fig. \ref{fig1:flowchart}, 3)
    \begin{enumerate}[leftmargin=0.1em]
        \item At time step $t$, compute the vector $\Delta = \vec{spins} \bullet Adj $, representing the change in energy if each individual node’s spin were flipped $(\Delta_i)$. This computation is from measuring the currents out of the crossbar columns when running inference, and subtracting pairwise columns. 
        \item Compare each $\Delta_i$ against a dynamic threshold $q$, derived from prior iteration statistics, and a control function $F$. $F$ is either $max(x)$ or $min(x)$ based on the problem. From this evaluation, the algorithm adaptively selects between a greedy approach and a simulated annealing approach.
        \item Identify all nodes with $\Delta_i < q$, and select up to $k$ of these nodes for flipping in the current iteration. Each flip is recorded both in the Spin Unit (within the Python simulator) and programmed back to the $ReRAM$ array on the hardware.
        \item Repeat the process until the termination condition is met. When no $\Delta_i$ in an iteration falls below the threshold for that cycle, terminate.
    \end{enumerate}
    \item \textbf{Output and Reporting}: Upon convergence, the algorithm returns the final spin vector from the Spin Unit, along with relevant metrics used for analyzing algorithmic efficiency and hardware performance.
\end{enumerate}

\begin{figure}
    \centering
    \includegraphics[width=0.9\linewidth]{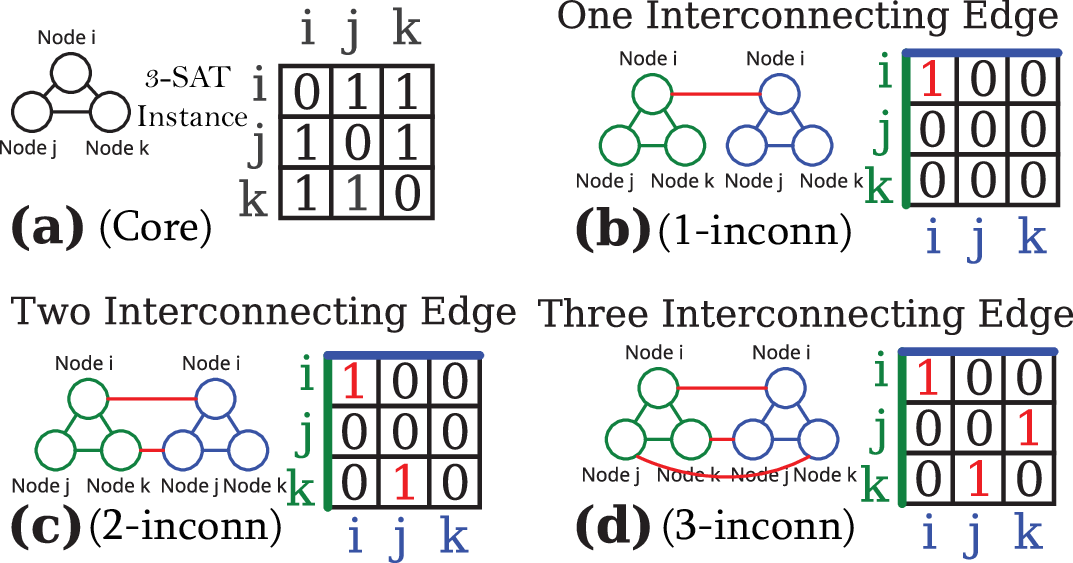}
    \includegraphics[width=1\linewidth]{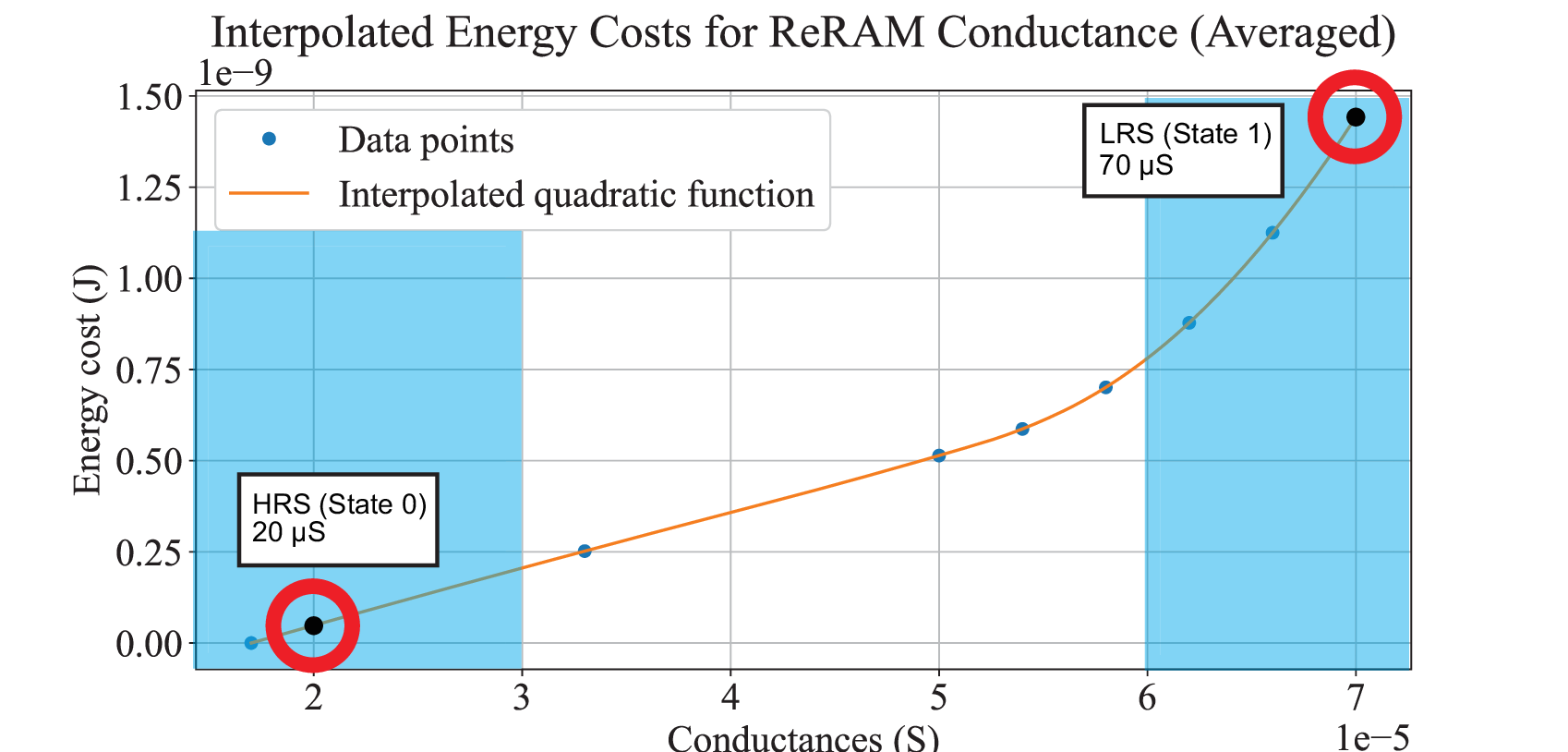}
    \caption{Key subgraph structures, and the average profiled energy states for the Skywater ReRAM. Red circles highlight the chosen HRS and LRS elected to be State 0 and State 1, respectively. Blue bars represent the error windows accepted.}
    \label{fig2:Subgraphs_energy}
    \vspace{-8mm}
\end{figure}

\begin{figure*}[h]
    \centering
    \includegraphics[width=0.9\linewidth]{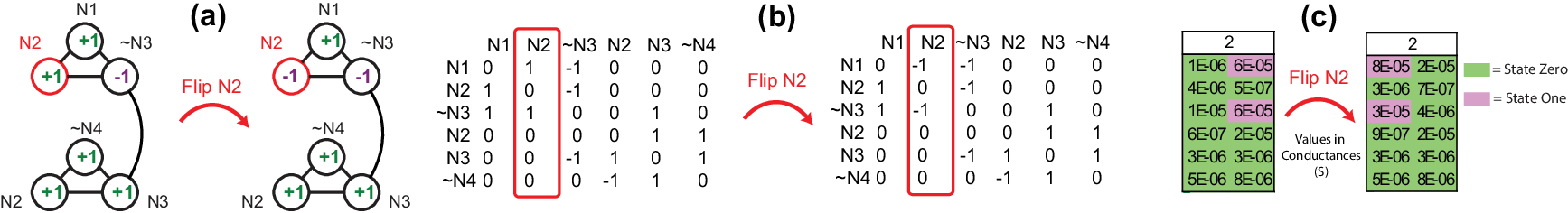}
    \caption{An example of the transformation of the Conductance in the ReRAM array between loop iterations.}
    \label{fig4:array}
    \vspace{-9mm}
\end{figure*}

\subsection{Physical ReRAM Fabrication and Characterization}

The ReRAM devices were fabricated in the SkyWater 130 nm CMOS process and integrated into a chip containing a fully addressable 32 × 16 ReRAM array with row and column decoders enabling cell-level programming and readout. Each cell uses a 2T–1R topology to provide precise write control and minimize leakage, and the chip is mounted on a PCB with I/O access for automated testing using a waveform generator and a Source Measure Unit (SMU). This measurement setup was used to comprehensively profile the devices’ conductance states, power consumption, and retention behavior, with the proposed algorithm in section \ref{SubSec:Algo} operating at 1-bit precision. 
Energy consumption was characterized by measuring the energy required to program cells from high- to low-conductance states. The total switching energy is defined as the absolute difference in stored energy between initial and final states. Fig. \ref{fig2:Subgraphs_energy} contains the complete interpolated energy graph for the average absolute energy differences for programming the ReRAM, enabling accurate estimation of transition energy across all allowed tolerance ranges. The work uses State 0 (20 $\mu$S) and State 1 (70 $\mu$S) highlighted by the red circles. The accepted tolerance window of 10 $\mu$S, bounded by the blue boxes, retains sufficient noise margin. The devices were both profiled on their worst case performance, providing an upper bound, as well as their average performance. Inference energy is computed from the applied voltage, measured current, and operation time; a detailed discussion of this profiling methodology is provided in \cite{Didin_2025}. 

\subsection{Simulated Supported Units} 

Operations external to the SkyWater ReRAM array are modeled in a Python-based simulator. This includes steps 1 and 2 of the algorithm, and implementing the q-unit, threshold unit, spin unit, and column-wise current summer/difference units. The q-unit and threshold units act as control modules governing algorithm dynamics, particularly for multi-tile operation, while the spin unit provides secondary storage for node spin values in the 3-SAT graph. This separation simplifies crossbar inference by decoupling spin state storage from candidate flipped states represented in the ReRAM array. In a full-chip implementation, the current summer/difference unit would be integrated with the array; in the experimental setup, individual cell currents are instead measured and aggregated in the simulator to enable fine-grained control and analysis.

\section{Benchmarks and Results}

Our results are presented in two parts. First, we analyze the power costs of the subarray structures in Fig. \ref{fig2:Subgraphs_energy}. Second, we evaluate the accuracy of ReRAM state transitions across iterations and relate this accuracy to the energy required to resolve each CNF instance.

\subsection{Key Graph Kernels: Benchmark} \label{subsec:KeyGraph}

In a given 3-SAT problem, each clause requires all three of its variables (nodes) to be mutually connected. Since these nodes are neighbors in the corresponding graph representation, they naturally form a Core kernel structure, as illustrated in Fig. \ref{fig2:Subgraphs_energy} $a$. This Core structure recurs along the diagonal of the adjacency matrix, thereby inducing repeated substructures within the associated ReRAM matrix.

A fundamental property of interest in the 3-SAT solving space is the \emph{density} of the problem \cite{coarfa2000random}, defined as the ratio of the number of clauses $(m)$ to the number of variables $(n)$, i.e., $m/n$. As the number of clauses increases while the number of variables remains fixed, the number of inter-clause connections necessarily grows (unless additional clauses are vacuously satisfiable and disconnected from all others). Studies indicate that around a density of $3.8$--$4.2$, 3-SAT instances undergo a phase transition that significantly increases their solving difficulty \cite{ercsey2011optimization}. Accordingly, we also examine the subgraph structures associated with inter-clause connections (inconn).

From Fig. \ref{fig2:Subgraphs_energy}, we identify three key interconnection patterns: clauses may be connected by one, two, or three edges; zero connections introduce no switching costs for column writing. While a node from one clause could simultaneously connect to two nodes from another clause, this scenario would require those nodes to represent the same literal from the original 3-SAT instance, reducing to a 2-SAT clause. Since our focus is on 3-SAT instances, we restrict attention to the aforementioned structures. This observation motivates the development of metrics for operations on these fundamental substructures, which could help extrapolate costs associated with larger arrays implemented using the same ReRAM devices.

\subsection{Key Graph Kernels: Results} \label{sec:graphKernelResults}

The first component of power consumption arises from initializing the ReRAM array. Starting from all cells in a low-conductance state, we recorded the energy required to initialize each core substructure from Fig. \ref{fig2:Subgraphs_energy}. Although substructures can be permuted through row or column operations, the total number of cells that must be set to high conductance remains constant. This also applies to any random initial spin assignment, where the only difference is which pos/neg representative column is programmed. The first section of Tab. \ref{tab2:results1} shows the average energy measured across 10 iterations for writing each substructure from all low conductance state. From Fig. \ref{fig2:Subgraphs_energy}, we see that the number of ReRAM that must be programmed to a high conducting state increases as a larger number of interconnections are represented. Subarray $a$ represents the most interconnections, where $b$ represents the fewest, and we see the same pattern within the results of Tab. \ref{tab2:results1}. A critical observation to note is that the differences between each subarray is not the aggregate of the additional ReRAM cells that are written. Referencing the average energy case in Fig. \ref{fig2:Subgraphs_energy}, it would be approximately $2.8\;nJ$ of energy increase per additional connection. However, we see only an increase of $0.7-2.4\;nJ$ between the increases of subarray $b\rightarrow c$ or $c\rightarrow d$ respectively. From subarray $d$ to subarray $a$, which is an increase in 3 more necessary interconnections, the increase is only $2.6\;nJ$ on average. 

The second component of power consumption arises during algorithm execution. We focus on the costs associated with flipping a column each iteration. Each column in the matrix contains one column from structure $a$ (the clause on the diagonal) along with one or more columns from structures $b$, $c$, or $d$. Since individual columns from $b$, $c$, and $d$ contain at most one adjacency, we separate measurements as shown in the second section of Table \ref{tab2:results1}. Flipping elements from $a$ consumes roughly twice the energy of $b,c,d$, consistent with $a$ columns representing twice as many adjacencies. We still note that the overall costs are below the direct writing costs for each ReRAM. Flipping a column of $b, c, d$ involves one pair of ReRAM flips, being $2.8 \;nJ$ again in the average case. This is seen at one standard deviation above the average energy for a column flip in $b, c, d$. The costs of flipping pairs of ReRAMs performs better than summing ReRAMs in isolation. 

The observed behaviors support exploring this implementation. By using only two state configuration, and allowing the 10 $\mu$S window, the costs associated with transitioning between high conductance and low conductance states have the opportunity to 'shortcut'. Any cells that go from the low bound of State 1, and stop at the high bound of State 0, inherently save energy costs in writing. This could be explored with allowing larger windows, or moving the accepted states closer together. In addition, allowing multiple 'in-between' transition states could further allow smaller energy writing stop gaps between switching back and forth. 

\begin{table}[]
\centering
\caption{Energy costs for programming and column switching of the fundamental matrices from Fig.~\ref{fig2:Subgraphs_energy}. Each entry is the average and standard deviation.}
\label{tab2:results1}
\begin{tabular}{c c c c}
\toprule
\makecell{\textbf{Subgraph}\\\textbf{Eng.}} & 
\makecell{\textbf{Array}\\\textbf{Type}} & 
\makecell{\textbf{Energy Costs}\\\textbf{Upper Bound}\\\textbf{(nJ)}} & 
\makecell{\textbf{Energy Costs}\\\textbf{Average Case}\\\textbf{(nJ)}} \\
\midrule

\multirow{4}{*}{Initialize} 
 & Core (a) & 21.2 $\pm$ 2.36 & 6.23 $\pm$ 0.893\\
 & 1-inconn (b) & 2.47 $\pm$ 0.41 & 0.574 $\pm$ 0.102\\
 & 2-inconn (c) & 4.51 $\pm$ 0.93 & 1.21 $\pm$ 0.313 \\
 & 3-inconn (d) & 12.2 $\pm$ 2.60 & 3.60 $\pm$ 1.12\\
\midrule

Program & Core (a) & 13.2 $\pm$ 2.28 & 3.85 $\pm$ 0.741 \\
Iteration & 1/2/3-inconn (b,c,d) & 6.79 $\pm$ 1.73 & 2.07 $\pm$ 0.619\\
\bottomrule
\end{tabular}
\vspace{-7mm}
\end{table}

\subsection{Full CNF Solving: Benchmark}

The ReRAM crossbar was profiled using the following small 3-SAT problems expressed in conjunctive normal form (CNF), note that $\vee = or$, $\wedge = and$. Each CNF is labeled with \#-X which represents the number of interconnections that appear in that CNF problem across the two clauses:
\vspace{-4mm}
\begin{enumerate}
\begin{multicols}{2}
\footnotesize	
    \item[0-X:] $(1 \vee 2 \vee 3) \wedge (4 \vee 5 \vee -6)$
    \item[1-X:] $(1 \vee -2 \vee 3) \wedge (2 \vee 3 \vee 4)$
    \item[2-X:] $(1 \vee 2 \vee -3) \wedge (-2 \vee 3 \vee -4)$
    \item[3-X:] $(1 \vee 2 \vee 3) \wedge (-1 \vee -2 \vee -3)$
\end{multicols}
\end{enumerate}

\vspace{-4mm}
These small test instances are designed to capture the core interactions between pairs of 3-SAT clauses as they would appear in larger benchmark sets, such as SATLIB \cite{satlib2011}. The problems were designed to remain small in order to effectively fit entirely on the Skywater 130nm ReRAM device. This provides complete power profiling by the physical devices, without any simulated devices that would be needed for larger problems.

\subsection{Full CNF Solving: Results}

During the main loop, the ReRAM cells, representing the graph nodes whose spins are flipped, need to be reprogrammed. This occurs every iteration, and we assess the ReRAM devices for storing new conductances and preserving previous ones. For example, Fig. \ref{fig4:array}$c$ shows $100\%$ accuracy in writing to cells, with neighboring cells flipping from $(0, 1)$ to $(1, 0)$ (rows 1 and 3), and all other rows remaining the same state. The combined accuracy metric is shown in column 2 of Tab. \ref{tab3:results2}. For each CNF problem, 10 tests were conducted with 10 iterations of the main loop, totaling 400 iterations. There was an aggregate accuracy of $91.0\%$. The algorithm tolerates ReRAM cell inaccuracies. At step $t$, the candidate nodes for flipping are determined by the threshold $q$ and $\Delta_i$. As example, if row 3 in Fig. \ref{fig4:array}$c$ is incorrectly written as $(1, 1)$, or $(0, 0)$, the effect on $\Delta_2$ is nullified. If row 1 is correct, the algorithm recognizes the node's behavior in flipping state with the remaining accurate connection. Only if row 3 is incorrectly written as $(0, 1)$, would that conflict with the column behavior. We conclude only pairwise errors along the same column could a node be improperly excluded/included as a candidate. The algorithm even still maintains accuracy if the $k$ nodes selected are a subset of the ideal nodes, even with pairwise write errors. This was supported with $92.5\%$ accuracy in determining the satisfiability of the CNFs.

\begingroup
\setlength{\tabcolsep}{3.2pt}   
\renewcommand{\arraystretch}{1.02} 
\setlength{\aboverulesep}{0.2ex}  
\setlength{\belowrulesep}{0.2ex}

\begin{table}[t]
\centering
\caption{Accuracy of ReRAM iterations and energy costs for each CNF problem. The last three columns are averages across 5 tests.}
\label{tab3:results2}
\begin{tabular*}{\columnwidth}{@{\extracolsep{\fill}} c c c c c}
\toprule
\makecell{\textbf{Iteration}\\\textbf{Var.}} &
\makecell{\textbf{ReRAM}\\\textbf{Iteration}\\\textbf{Acc.}} &
\makecell{\textbf{Execute}\\\textbf{Energy}\\\textbf{Upper} \\\textbf{Bound (nJ)}} &
\makecell{\textbf{Execute}\\\textbf{Energy}\\\textbf{Average} \\\textbf{Case (nJ)}} &
\makecell{\textbf{Avg. Inference}\\\textbf{Energy (nJ)}} \\
\midrule
CNF 0-X & 94.3\% & 128 & 37.5 & 1270 \\
CNF 1-X & 87.6\% & 137 & 40.7 & 5840 \\
CNF 2-X & 89.9\% & 151 & 43.7 & 4230 \\
CNF 3-X & 92.1\% & 189 & 56.8 & 2250 \\
\midrule
Overall & 91.0\% & 151 & 44.7 & 3400 \\
\bottomrule
\end{tabular*}

\vspace{-7mm}
\end{table}

\endgroup

Finally, we profiled the energy costs for solving each CNF instance, shown in columns 3, 4, and 5 of Tab. \ref{tab3:results2}. The inference energy costs result from storing the ReRAM array each loop for data completeness, and would not be reflected in proper deployment. We observe a correlation between the execute energy and the density of the CNF problem. CNF 3-X has maximum interconnections and used the most energy, compared to CNF 0-X which has the minimum interconnections and used the least. We compared the profiled energy values with the results in Tab. \ref{tab2:results1}. If the power contributions in the crossbar array added linearly, CNF 0-X — which writes two instances of subgraph $a$ and iteratively swaps a column containing only that subgraph —would consume approximately $50.96 \;nJ$ in the average case for array writing and programming. However, comparison with Column 4 of Tab. \ref{tab3:results2} shows a lower measured value of $37.5 \; nJ$, indicating that total energy does not scale additively. This result suggests that meaningful energy savings can be achieved through this accelerated approach. 

\section{Conclusion}

Efficiently solving 3-SAT problems remains a key challenge for computational workloads. A critical factor in this efficiency is the power cost of computation, combined with power costs of memory transfer. In this work, we demonstrated that our ReRAM crossbar implementation achieves low energy consumption while maintaining high memory accuracy between iterations. By decomposing 3-SAT problems into fundamental kernel operations, we provide a benchmarked framework that can guide the design and development of larger, scalable devices capable of addressing the substantial energy demands of these computations. 

\bibliographystyle{IEEEtran}
\bibliography{references2.bib}

\end{document}